\documentclass{PoS}

\usepackage{epsfig}

\PoS{PoS(LAT2005)338}

\title{Calculating $B_K$ using a mixed action }

\ShortTitle{Calculating $B_K$ using a mixed action}

\author{\speaker{Jongjeong Kim} \\ Department of Physics, Seoul
National University, Seoul, 151-747, South Korea \\ E-mail:
\email{rvanguard@phya.snu.ac.kr}}

\author{Taegil Bae \\ Department of Physics, Seoul National
University, Seoul, 151-747, South Korea \\ E-mail:
\email{esrevinu@phya.snu.ac.kr}}

\author{Weonjong Lee \thanks{This research is supported in part by the
KOSEF grant (R01-2003-000-10229-0), by the MOST/KISTEP grant of
international collaboration, and by the BK21 program.}\\ Center for
Theoretical Physics, Department of Physics, Seoul National University,
Seoul, 151-747, South Korea \\ E-mail: \email{wlee@phya.snu.ac.kr}}

\abstract{ We present preliminary results of $B_K$ calculated using
improved staggered fermions with the mixed action (valence quarks =
HYP staggered fermions and sea quarks = AsqTad staggered fermions). We
analyze the data based upon the prediction by Van de Water and
Sharpe. A hint of consistency with the prediction is observed. We also
present preliminary results of $B_8^{(3/2)}$ and $B_7^{(3/2)}$.}

\FullConference{XXIIIrd International Symposium on Lattice Field Theory\\
25-30 July 2005\\
Trinity College, Dublin, Ireland}

\begin{document}
\section{$B_K$}
\label{sec:bk}
\begin{figure}[h!]
\epsfig{file=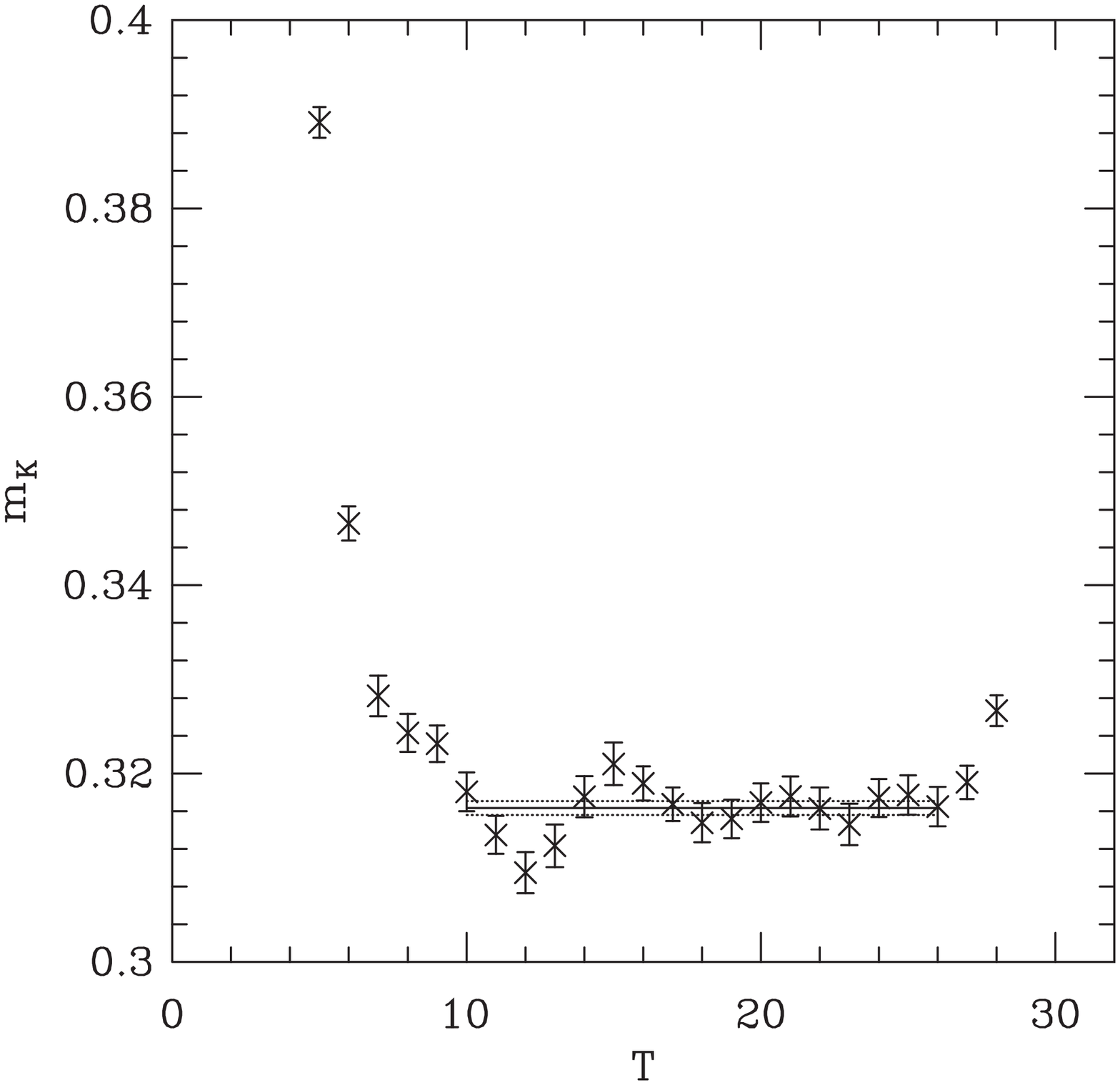, width=0.5\textwidth}
\epsfig{file=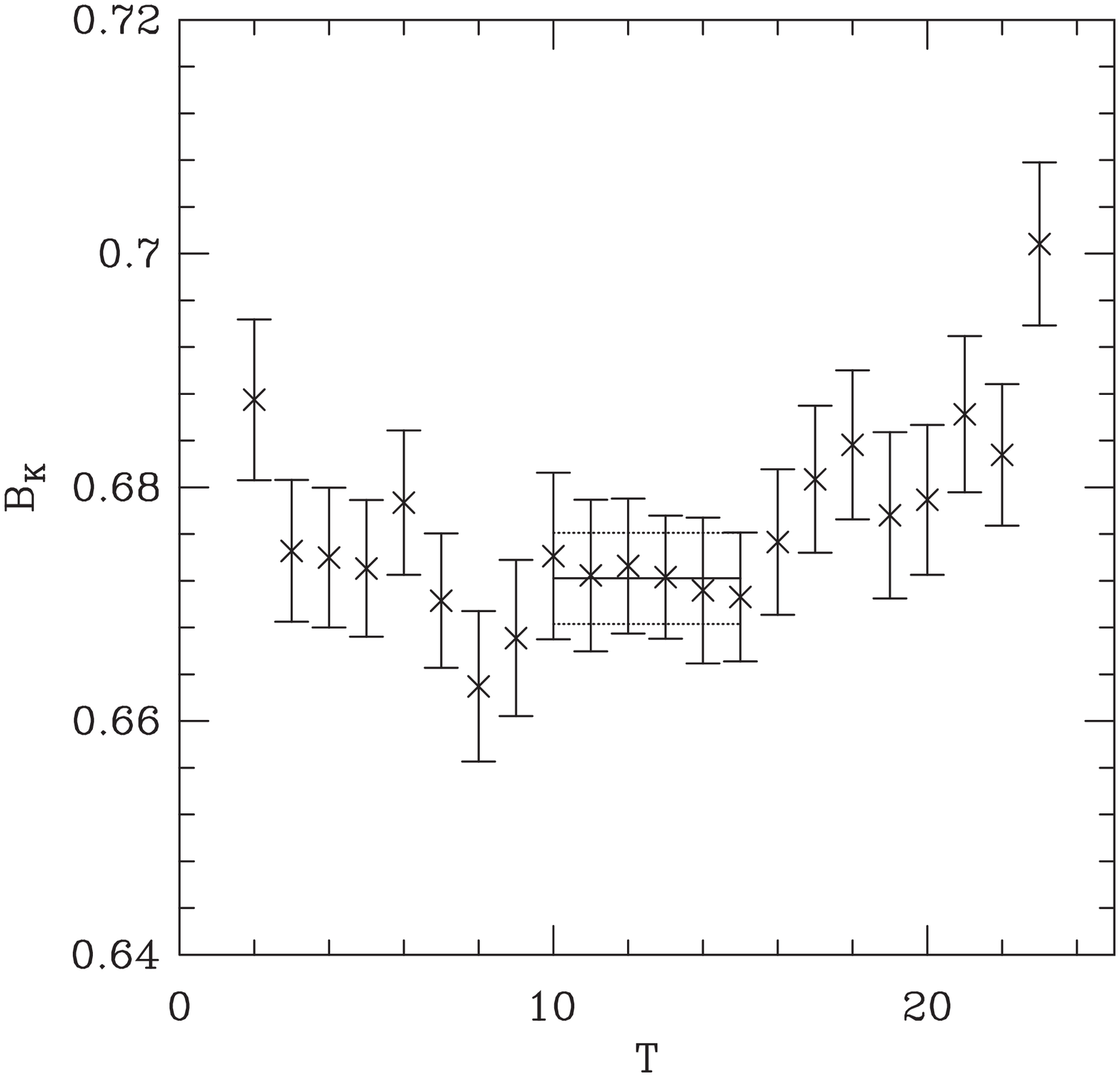, width=0.5\textwidth}
\caption{$m_K$ vs. T (left) and $B_K$ vs. T (right)
at $m_s = m_d = 0.03$}
\label{fig:mpi-t:bk-t}
\end{figure}

The size of indirect CP violation in the neutral Kaon system is, in
experiment, parameterized by $\epsilon$, which is proportional to the
kaon bag parameter $B_K$ defined as
\begin{eqnarray}
   B_K &=& \frac{\langle \bar{K}_0 | [\bar{s} \gamma_\mu (1-\gamma_5) d]
      [\bar{s} \gamma_\mu (1-\gamma_5) d] | K_0 \rangle }{
      \frac{8}{3} \langle \bar{K}_0 | \bar{s} \gamma_\mu\gamma_5 d | 0 \rangle
      \langle 0 | \bar{s} \gamma_\mu\gamma_5 d | K_0 \rangle }
\end{eqnarray}
A precise determination of $B_K$ constrains the CKM matrix, which might
lead us to a window of new physics beyond the standard model.
Hence, there have been a number of efforts to calculate $B_K$ with
higher precision.
As pointed out in Ref.~\cite{ref:wlee:1}, the large scaling violation
observed in the calculation using unimproved staggered fermions can be
reduced remarkably by improving staggered fermions with HYP fat links.
Even though the scaling violation is taken care of by improving
staggered fermions, there has been an uncertainty originating from the
quenched approximation.
Hence, it has been crucial to perform a numerical study in unquenched
QCD so that we can remove the uncertainty due to quenched
approximation.
In this paper, we describe our first attempt to perform a numerical
study on $B_K$ in partially quenched QCD, while minimizing the scaling
violation using improved staggered fermions.
We use HYP staggered fermions as valence quarks and AsqTad staggered
fermions as sea quarks (we call this a ``mixed action'').
Details of the simulation parameters are summarized in
Table~\ref{tab:param}.
\begin{table}[h!]
\begin{center}
\begin{tabular}{c | c}
\hline
parameter & value \\
\hline
sea quark & AsqTad Staggered \\
valence quark & HYP Staggered \\
$\beta $ & 6.76 ($N_f=2+1$ QCD) \\
\# of confs & 593 \\
lattice & $20^3 \times 64$ \\
sea quark masses & $m_{u,d} = 0.01$, $m_s =0.05$ \\
valence quark mass & 0.01, 0.02, 0.03, 0.04, 0.05 \\
\hline
\end{tabular}
\end{center}
\caption{Parameters for the numerical study}
\label{tab:param}
\end{table}
We measure weak matrix elements and hadron spectrum over a subset of
the MILC gauge configurations \cite{ref:milc:1}.
We set up U(1) noise sources at $T=0$ and $T=26$, which project out
only pseudo-Goldstone pions ($\gamma_5 \otimes \xi_5$) and exclude all
the other non-Goldstone pions.
The kaon signal as a function of Euclidean time $T$ is shown in the
lefthand side of Fig.~\ref{fig:mpi-t:bk-t}.
Here, we observe that there is a noticeable contamination from excited
states in the range of $0 \le T < 10$.
Hence, the best fitting range for $B_K$ is $ 10 \le t \le 15 $ in
order to exclude the contamination from excited states, as you can see
in the righthand side of Fig.~\ref{fig:mpi-t:bk-t}.

Ref.~\cite{ref:damir:1} presents a chiral behavior of $B_K$ and its
finite volume effect in the case of $N_f=2$ partially quenched QCD.
Although this result is interesting, it does not apply directly to our
numerical study mainly because it is in $N_f = 2+1$ partially quenched
QCD.
Recently, Van de Water and Sharpe have calculated the chiral behavior
of $B_K$ in $N_f=2+1$ partially quenched QCD using staggered chiral
perturbation theory \cite{ref:sharpe:1}.
The result for the degenerate valence quark mass combination ($m_x =
m_y$) is
\begin{eqnarray}
B_K &=& \tilde{c}_0 \bigg( 1 +
\frac{1}{48\pi^2 f^2} \Big[ I_{\rm conn} + I_{\rm disc} 
+ \tilde{c}_1 m_{xy}^2 
+ \tilde{c}_3 ( 2 m_U^2 + m_S^2 ) \Big]
\bigg)
\label{eq:bk:fit}
\\
I_{\rm conn} &=& 6 m_{xy}^2 \tilde{l}(m_{xy}^2) 
- 12 l(m_{xy}^2)
\\
I_{\rm disc} &=& 0
\\
l(X) &=& X \log( X/ \Lambda^2 ) + {\rm F.V.}
\\
\tilde{l}(X) &=& - [ \log( X / \Lambda^2 ) + 1] + {\rm F.V.} 
\end{eqnarray}
where $f = 132$ MeV, $\tilde{c}_i$ are unknown dimensionless
low-energy constants and ${\rm F.V.}$ represents a finite volume
correction.
The notations for the various meson masses are as follows for those
composed of sea quarks:
\begin{eqnarray}
m_U^2 = 2 \mu m_d, \qquad 
m_S^2 = 2 \mu m_s, \qquad
m_\eta^2 = (m_U^2 + 2 m_S^2) / 3
\end{eqnarray}
and as follows for those composed of valence quarks
\begin{eqnarray}
m_X^2 = 2 \mu m_x, \qquad 
m_Y^2 = 2 \mu m_y, \qquad
m_{xy}^2 = \mu (m_x + m_y)
\end{eqnarray}
Here, we set sea quark masses to $m_u=m_d \ne m_s$ and
the two valence quark masses are $m_x$ and $m_y$. 
In this paper, we consider only the case of $m_x=m_y$
({\em i.e.} $m_X^2 = m_Y^2 = m_{xy}^2$).
\begin{figure}[h!]
\begin{center}
\epsfig{file=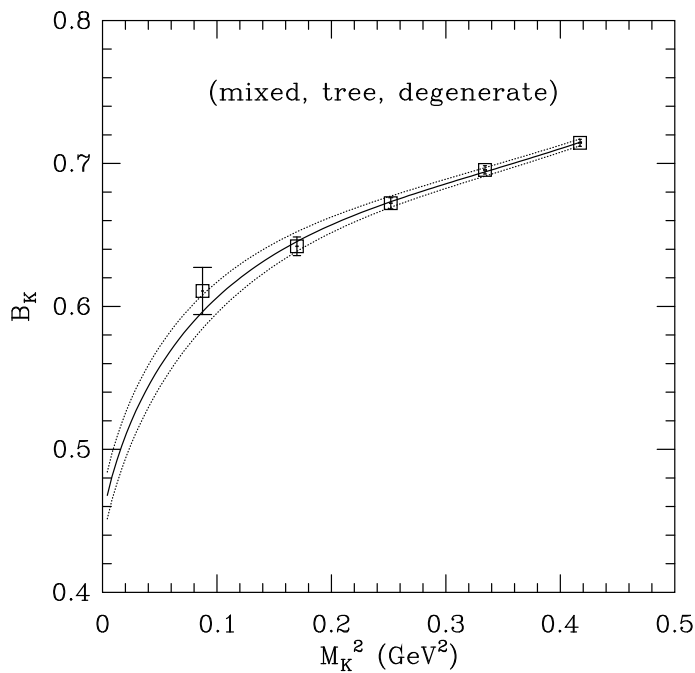, width=0.6\textwidth}
\end{center}
\caption{$B_K$ vs. $M_K^2$}
\label{fig:bk}
\end{figure}
In Fig.~\ref{fig:bk}, we plot $B_K$ data as a function of kaon mass
squared $M_K^2$.
Here, all of them have degenerate valence quarks ($m_x = m_y$).
We fit the data to the form of Eq.~(\ref{eq:bk:fit}) suggested by
chiral perturbation theory:
\begin{eqnarray}
B_K &=& c_1 \bigg( 1 +
\frac{1}{48\pi^2 f^2} \Big[ I_{\rm conn} + I_{\rm disc} \Big] \bigg)
+ c_2 m_{xy}^2 + c_4 m_{xy}^4
\end{eqnarray}
where the cut-off scale $\Lambda$ is set to $\Lambda=4 \pi f$ and
the remaining scale dependence is absorbed into $c_2$.
The fitting results are summarized into Table \ref{tab:bk:fit}.

In Table \ref{tab:bk:fit}, the $\chi^2$ is rather high when we take
into account the fact that all the data are correlated.
In fact, one can observe that the fitting curve does not fit the
lightest two data points very well in Fig.~\ref{fig:bk}.
In other words, the fitting curve miss the lightest two data points in
the opposite direction.
In Ref.~\cite{ref:sharpe:1}, it is pointed out that the contribution
from the non-Goldstone pions are so significant that the curvature of
the fitting curve becomes smoother, which is consistent with what we
observe in Fig.~\ref{fig:bk}.
However, the full prediction from staggered chiral perturbation
contains 21 unknown low-energy constants for a single lattice spacing
and 37 unknown low-energy constants for the full analysis
\cite{ref:sharpe:1}.
In order to determine all of them, it is required to carry out a
significantly more extensive numerical work including data with
non-degenerate quarks \cite{ref:wlee:2}.
\begin{table}[h!]
\begin{center}
\begin{tabular}{| c | c | c |}
\hline
parameters & average & error \\
\hline
$c_1$           & 0.4488    & 0.0162 \\
$c_2$           & $-$1.4883 & 0.1883 \\
$c_3$           & $-$       & $-$ \\
$c_4$           & 1.4533    & 0.2207 \\
\hline
$\chi^2$/d.o.f. & 0.6812    & 0.6 \\
\hline
\end{tabular}
\end{center}
\caption{Fitting results for $B_K$}
\label{tab:bk:fit}
\end{table}

In the current analysis of $B_K$ data, we match the lattice results to
the continuum values at the tree level.
In this respect, the results are preliminary.
Hence, we plan to calculate the one-loop corrections to the four
fermion operators in near future.
In addition, note that there exists a different approach to $B_K$
using staggered fermions \cite{ref:gamiz:1}.

\section{$B_7^{(3/2)}$ and $B_8^{(3/2)}$}
\label{sec:b7:b8}
In the case of direct CP violation $\epsilon'/\epsilon$, there are two
contributions which interfere with each other destructively: the
$\Delta I =1/2$ part and $\Delta I = 3/2$ part.
The $\Delta I = 3/2$ part is dominated by the electroweak penguin
contribution from $Q_8$ and $Q_7$.
In this section, we focus on the bag parameters of $Q_8$ and $Q_7$.

\begin{figure}[h!]
\epsfig{file=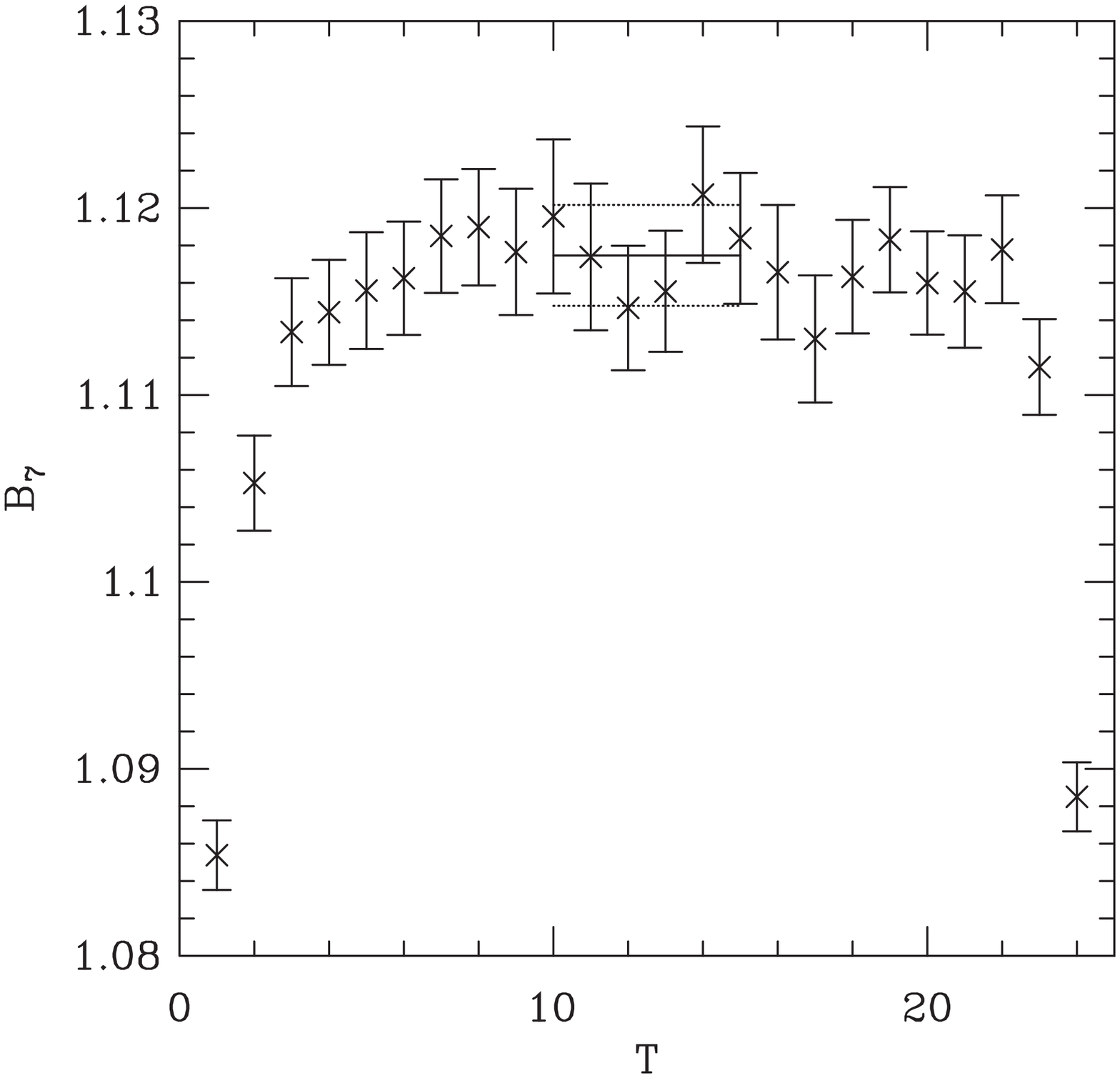, width=0.5\textwidth}
\epsfig{file=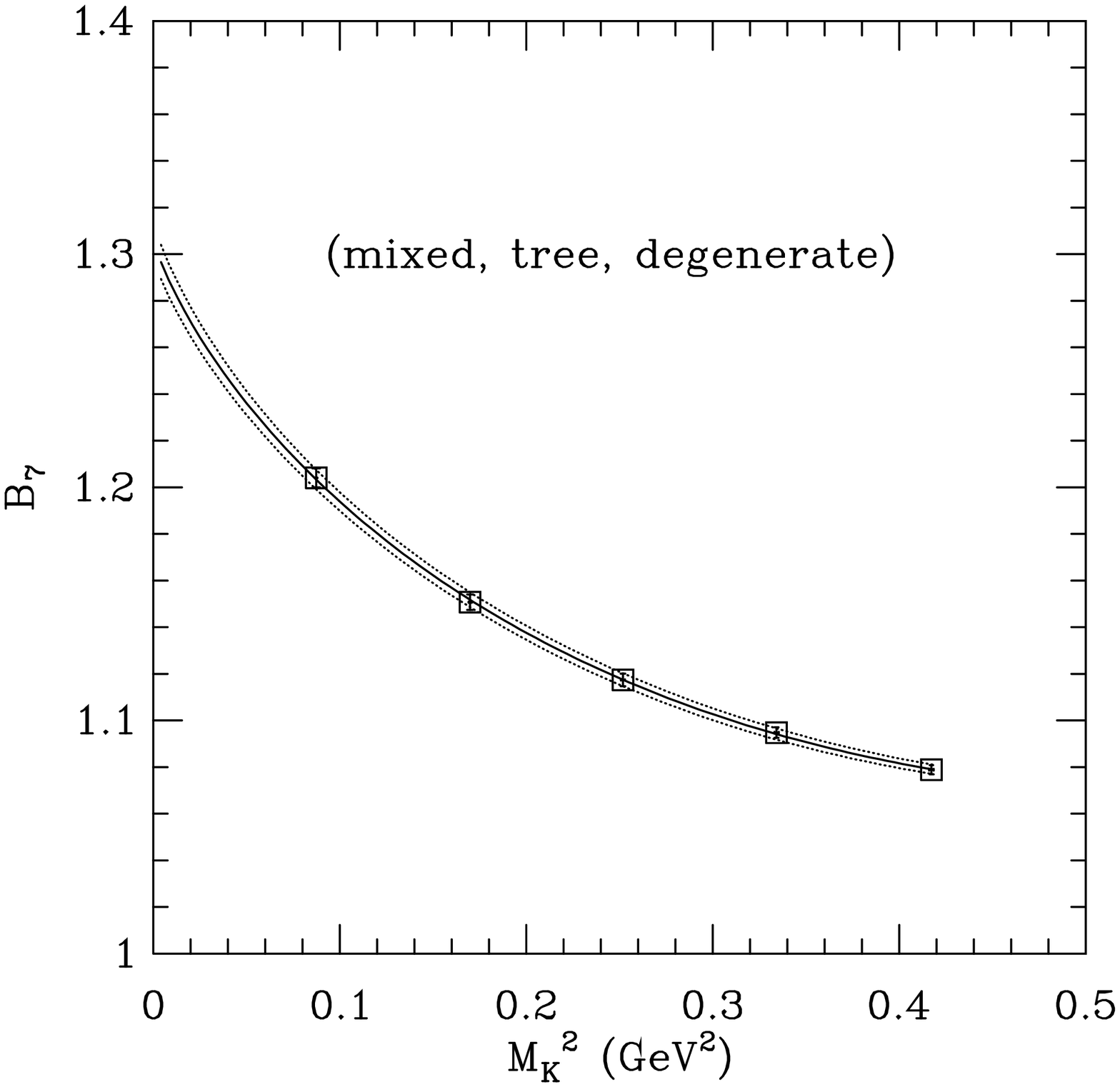, width=0.5\textwidth}
\caption{$B_7$ vs. T at $m_s = m_d = 0.03$ (left) and $B_7$
vs. $M_K^2$ (right)}
\label{fig:b7}
\end{figure}
The $B_7^{(3/2)}$ data as a function of Euclidean time T at quark mass
$m_x=0.03$ is presented in the lefthand side of Fig.~\ref{fig:b7}.
The fitting range is the same as for $B_K$.
In the righthand side of Fig.~\ref{fig:b7}, $B_7^{(3/2)}$ data is
plotted as a function of $M_K^2$.
We fit the data to the form suggested by chiral perturbation theory:
\begin{eqnarray}
B_{7,8}^{(3/2)} &=& c_1 + c_2 m_{xy}^2 + 
c_3 m_{xy}^2 \log( m_{xy}^2 / \Lambda^2 )
\label{eq:b7:fit}
\end{eqnarray}
The results are summarized in Table \ref{tab:b7:b8}.
Here, note that we match the lattice results to the continuum values
at the tree level.
In this respect, the results are preliminary.
\begin{table}[h!]
\begin{center}
\begin{tabular}{| c | c | c |}
\hline
parameters & $B_7^{(3/2)}$ & $B_8^{(3/2)}$ \\
\hline
$c_1$ & 1.3068 $\pm$ 0.0078    & 1.2505 $\pm$ 0.0074 \\
$c_2$ & $-$0.1889 $\pm$ 0.0093 & $-$0.1439 $\pm$ 0.0091 \\
$c_3$ & 0.4080 $\pm$ 0.0228    & 0.3175 $\pm$ 0.0222 \\
\hline
\end{tabular}
\end{center}
\caption{Fitting results for $B_7^{(3/2)}$ and $B_8^{(3/2)}$}
\label{tab:b7:b8}
\end{table}
\begin{figure}[h!]
\epsfig{file=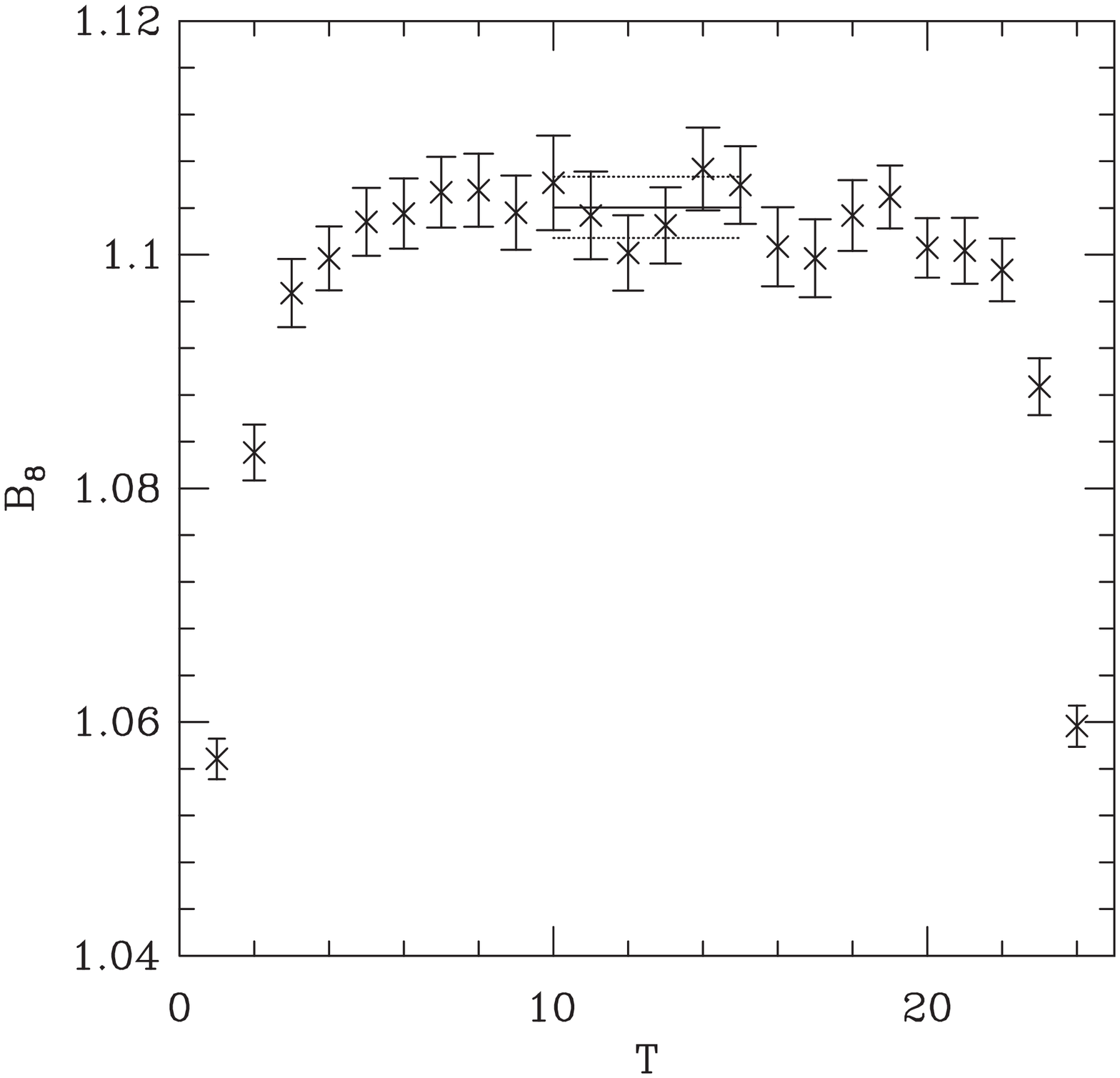, width=0.5\textwidth}
\epsfig{file=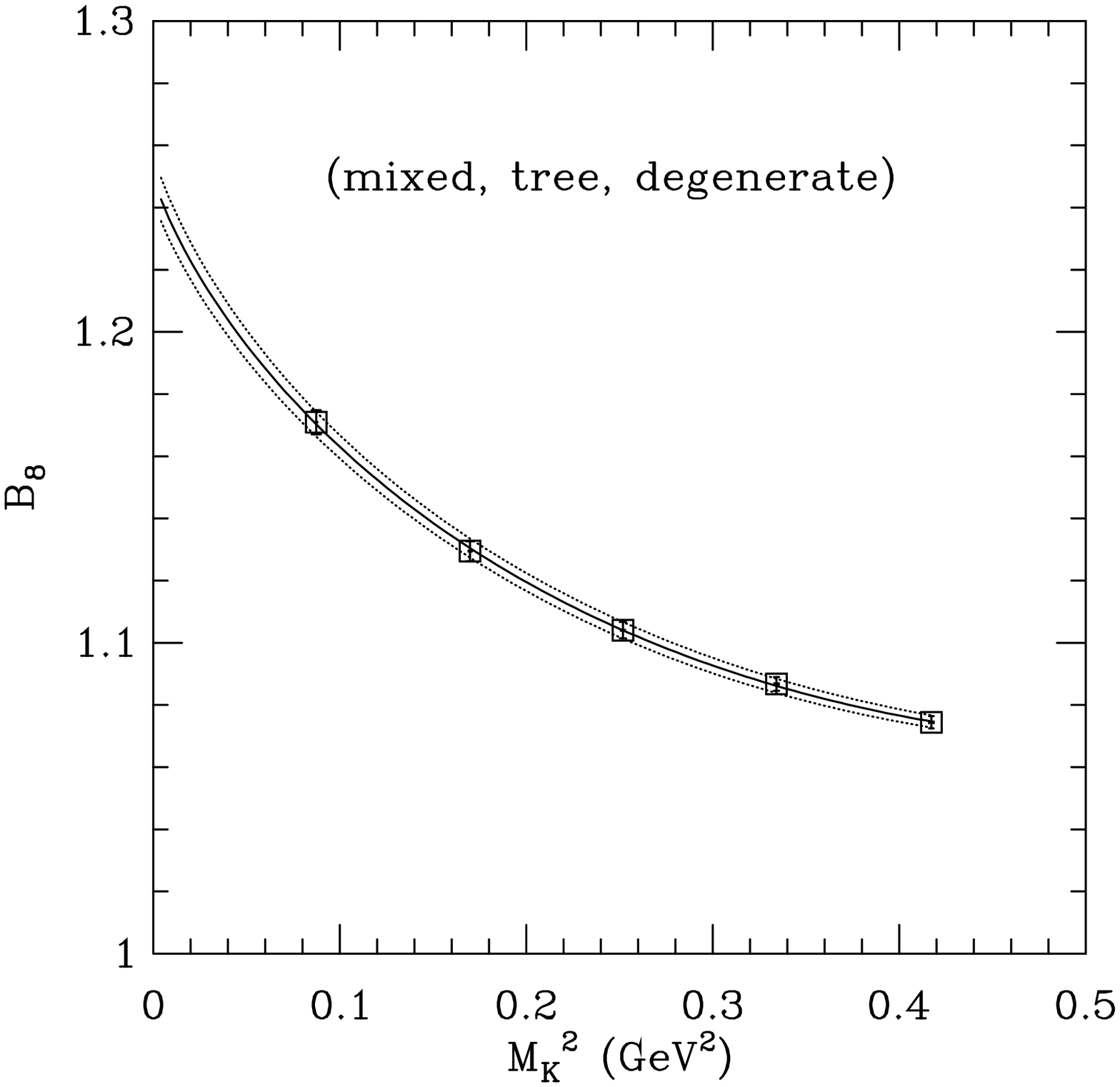, width=0.5\textwidth}
\caption{$B_8$ vs. T at $m_s = m_d = 0.03$ (left) and $B_8$
vs. $M_K^2$ (right)}
\label{fig:b8}
\end{figure}
We present $B_8^{(3/2)}$ data as a function of Euclidean time in the
lefthand side of Fig.~\ref{fig:b8}.
We plot $B_8^{(3/2)}$ as a function of $M_K^2$ in the righthand side of
Fig.~\ref{fig:b8}.
We fit the data to the form of Eq.~(\ref{eq:b7:fit}) and the
preliminary results are summarized in Table \ref{tab:b7:b8}.

In order to perform a data analysis with higher precision, we need to
know the chiral behavior of $B_8^{(3/2)}$ and $B_7^{(3/2)}$.
Hence, we plan to calculate this using staggered chiral perturbation
theory.


\begin{thebibliography}{99}
%
\bibitem{ref:wlee:1} Weonjong Lee {\em et al.}, \emph{Testing improved
staggered fermions with $m_s$ and $B_K$}, 
\emph{Phys.~Rev.} D{\bf 71} (2005) 094501,
[{\tt hep-lat/0409047}].
%
\bibitem{ref:milc:1} C. Aubin {\em et al.}, \emph{ Light hadrons with
improved staggered quarks: approaching the continuum limit },
\emph{Phys.~Rev.} D{\bf 70} (2004) 094505, [{\tt hep-lat/0402030}].
%
\bibitem{ref:damir:1} Damir Becirevic and Giovanni Villadoro, \emph{
Impact of the finite volume effects on the chiral behavior of $f_K$
and $B_K$ }, \emph{Phys.~Rev.} D{\bf 69} (2004) 054010, [{\tt
hep-lat/0311028}].
%
\bibitem{ref:sharpe:1} Ruth Van de Water and Stephen Sharpe, \emph{
$B_K$ in Staggered Chiral Perturbation Theory }, [{\tt
hep-lat/0507012}].
%
\bibitem{ref:wlee:2} Taegil Bae, Jongjeong Kim and Weonjong Lee,
\emph{Non-degenerate quark mass effect on $B_K$ with a mixed action},
in proceedings of \emph{XXIIIrd International Symposium on Lattice
Field Theory}, PoS (LAT2005) 335, [{\tt hep-lat/0510008}].
%
\bibitem{ref:gamiz:1} Elvira Gamiz {\em et al.},\emph{Dynamical
determination of $B_K$ from improved staggered quarks}, in proceedings
of \emph{XXIIIrd International Symposium on Lattice Field Theory}, PoS
(LAT2005) 347, [{\tt hep-lat/0509188}]
%
\end{thebibliography}
\end{document}